\documentclass[12pt]{article}
\usepackage[letterpaper, top=1.00in, bottom=1.00in, left=1.00in, right=1.00in]{geometry}
\usepackage{graphicx}
\usepackage{amsmath}

\newcommand{\bsigma}{\boldsymbol{\sigma}}

\begin{document}
\title{Chromatic polynomials of random graphs}
\author{Frank~Van~Bussel$^{1,2,3}$, Christoph~Ehrlich$^{4}$, Denny~Fliegner$^{1}$, Sebastian \\ Stolzenberg$^{5}$ and Marc~Timme$^{1,2,3}$
 \\[10pt]
\parbox{5.5in}{\it \small $^1$~Max Planck Institute for Dynamics and Self-Organization (MPIDS), G\"{o}ttingen, Germany; \; $^{2}$~Georg August University School of Science (GAUSS), G\"{o}ttingen, Germany; \; $^{3}$~Bernstein Center for Computational Neuroscience (BCCN) G\"{o}ttingen, Germany; \; $^{4}$~Department of Physics, Technical University of Dresden, Dresden, Germany; \; $^{5}$~Department of Physics, Ithaca \& Weill Cornell Medical College, New York City, Cornell University, NY, USA}
}
\date{December 21, 2009}

\maketitle

\begin{abstract}
Chromatic polynomials and related graph invariants are central objects
in both graph theory and statistical physics. Computational difficulties,
however, have so far restricted studies of such polynomials to graphs
that were either very small, very sparse, or highly structured. Recent
algorithmic advances (\emph{New~J.~Phys.}~11:023001, 2009) now
make it possible to compute chromatic polynomials for moderately
sized graphs of arbitrary structure and number of edges. Here we present
chromatic polynomials of ensembles of random
graphs with up to 30 vertices, over the entire range of edge density.
We specifically focus on the locations of the zeros of the polynomial
in the complex plane. The results indicate that the chromatic zeros
of random graphs have a very consistent layout. In particular, the
{\em crossing point}, the
point at which the chromatic zeros with non-zero imaginary part approach
the real axis, scales linearly with the average degree over most of
the density range. While the scaling laws obtained are purely empirical,
if they continue to hold in general there are significant implications:
the crossing points of chromatic zeros in the thermodynamic limit separate
systems with zero ground state entropy from systems with positive ground
state entropy, the latter an exception to the third law of thermodynamics.
\end{abstract}

\section{Background}
The chromatic polynomial was introduced by Birkhoff in 1912 \cite{birkhoff}
as a way to bring complex analysis to bear on
what was then still the 4-colour conjecture. It is closely related
to the Tutte polynomial and valuations of these polynomials provide
information about many important graph invariants \cite{tutte}.
In recent decades chromatic polynomials, and in particular their zero sets (an equivalent representation), have
become the focus of attention from physicists due to their connection
to the Potts model in statistical physics \cite{fywu}.
For specific lattices this has given us very detailed information about
chromatic zeros \cite{shrock01,sokal01,shrock97}; however,
there has been little progress towards understanding their location
for general graphs. This is in large part due to limited computational
accessibility; for all but the smallest, sparsest,
or most highly structured graphs calculating the chromatic polynomial
is extremely difficult because generally, the computation time increases
exponentially in the number of edges \cite{wilf}. Since
we lack instances of chromatic polynomials for `ordinary' graphs
from which to build intuition, we currently have little idea what
to expect for moderately sized graphs that are not extremely sparse or highly
structured. Hence we possess no basic background for comparison
of the few non-minimal instances known so far, cf. e.g. \cite{shrock98,sokal01}.

Random graphs constitute a natural first candidate towards
building such an intuition. Since random graphs serve as initial
approximations or null models of systems with unknown structure, they
are ubiquitous in scientific and mathematical research and commonly
used as testbeds for theories, algorithms, and analytic tools. Random
graph theory starts with the work of Erd\"{o}s and R\'{e}nyi in the late 1950's,
and the standard random graph (where each edge is present independently
with the same fixed probability) is therefore usually named
an Erd\"{o}s-R\'{e}nyi (ER) random graph. ER random graphs have been shown
to have some remarkably robust properties and precise invariants \cite{janson}.
This applies to several problems related to chromatic polynomials, such as the value of the
chromatic number of a random graph,
\begin{equation}
\left(\frac{1}{2}+o(1)\right) \log{\left(\frac{1}{1-p}\right)} \frac{N}{\log{N}},
\end{equation}
where $N \gg 1$ is the number of vertices and $p$ is the edge probability \cite{bollobas}. However, to the best
of our knowledge, the chromatic polynomial itself has not to date
been subject of such investigation. Given the overlapping interests
of random graph theory and statistical physics this might seem somewhat
surprising, but computational accessibility has so far been highly limited.

In this Communication, we start to fill this gap using a promising
vertex based, symbolic pattern matching method developed recently \cite{timme}
to compute chromatic polynomials of moderately sized graphs.
Like other techniques developed for physics applications (cf.~\cite{hartmann}) it is capable of
fast computations on graphs consisting of periodically repeating subgraphs;
however, one of the new method's important features is that it also works on arbitrary
graphs, for example on samples of various random-bond-diluted lattices
\cite{timme}. It performs significantly better than standard general
purpose methods, allowing access to larger graphs with an arbitrary number
of edges than have previously been considered,
such as three dimensional cubic lattices.

Here we compute and analyze chromatic polynomials for
moderately sized ER random graphs across the entire range of edge
densities for selected $N$. Our aim is to obtain general estimates
for chromatic zero locations with respect to both mean value and variability,
which can then be compared to what is known for specific lattices
and other graphs. Since for decent statistics chromatic polynomials
of many graphs for each given set of parameters need to be computed,
the maximum value we consider here is $N=30$; chromatic polynomials
for individual graphs with higher $N$ are possible to compute, but
resource constraints make computing a large number of realizations
problematic.

As indicated in the following, the chromatic zeros of random graphs are laid out in a very regular
way by density. While unanticipated, this feature is in fact entirely
consistent with previous random graph results; until now, however,
not enough was known for informed speculation on the subject. Based
on our data we provide a scaling law for the point at which the complex
zeros approach the real line, as well as estimates for various other
quantities of interest.

\section{Chromatic Polynomials, Partition Functions, and their zeros}
We begin with some notation: given a graph $G$ with vertex set
$V = \{v_1,v_2,\dots,v_N\}$ and edge set $E = \{e_1,e_2,\dots,e_M\}$, and a set of colours $C=\{1,2,\dots,q\}$,
we say that a {\em proper $q$ colouring} of $G$ is an assignment
of values from $C$ to the $v_i \in V$ such that no two vertices
connected by an edge share the same value. $G$ is {\em $q$-colourable}
if there is a proper colouring of $G$ using $q$ or fewer colours;
the {\em chromatic number} of $G$ is then the minimum $q$ such
that $G$ is $q$-colourable. The {\em chromatic polynomial} $P(G,q)$
is the associated counting function for proper $q$-colourings of
$G$; that is, it tells us how many ways we can colour $G$ with at
most $q$ colours.

The representation
\begin{equation} \label{CPdef}
P(G,q) = \sum_{\sigma_{N}=1}^{q} \cdots \sum_{\sigma_{1}=1}^{q} \, \prod_{(i,j) \in E} (1-\delta_{\sigma_{i}\sigma_{j}})
\end{equation}
of the chromatic polynomial in terms of sums over polynomials in Kronecker deltas
$\delta_{\sigma_{i}\sigma_{j}}$ seems particularly suited for computations
\cite{timme} and also directly shows a link to Potts partition functions in statistical physics
(see below). Here every $\bsigma=(\sigma_{1},\dots,\sigma_{N})$ is
an assignment of values (colors) from $\{1,2,...,q\}$ to the $N$
vertices of $G$ and the product runs over all edges $e = (i,j)$ of $G$.
For a given colouring $\bsigma$ the product equals one if no two
adjacent vertices have the same colour, and zero otherwise; it functions
as an indicator that the assignment is a proper colouring of $G$.
Equivalently, $\bsigma$ can be regarded as a global microscopic state
of an antiferromagnetic Potts model with the individual $\sigma_{i}$'s
being local states or spin values. From this viewpoint (\ref{CPdef})
counts the number of energy minimizing global states.

The connection to statistical physics arises from the fact that the
chromatic polynomial equals the antiferromagnetic Potts partition
function $Z(G,q,T)$ in the zero temperature limit, since $Z(G,q,T)$
counts the number of spin configurations where all neighboring
spins disalign \cite{fywu}. Let us be more specific: The standard $q$-state Potts
model was introduced in 1952 as a generalization of Ising's 2-state
model for interactions on a crystal lattice \cite{baxter,potts,onsager,fywu}.
It describes systems in which site variables (magnetic moments, spins
or other kinds of local states) can take one of $q$ different values,
and interactions occur only between neighbouring sites on a lattice
that are in the same state. The total energy in the global state $\bsigma$
is given by the Hamiltonian
\begin{equation} \label{eqpotts}
H(\bsigma) = -J \sum_{(i,j)\in E} \delta_{\sigma_{i}\sigma_{j}}
\end{equation}
where $J$ is the interaction strength, and $E$ is the edge set
of the underlying graph $G$. The partition function for this system
at positive temperature $T = (k_{B}\beta)^{-1}$ is
\begin{equation} \label{eqpartn}
Z(G,q,T) = \sum_{\bsigma}e^{-\beta H({\bsigma})} = \sum_{\bsigma} \prod_{(i,j) \in E} (1 + (e^{\beta J} - 1) \delta_{\sigma_{i}\sigma_{j}})
\end{equation}
where $k_{B}$ is the Boltzmann constant. If $J<0$ (antiferromagnet)
the $T \rightarrow 0$ limit $e^{\beta J} \rightarrow 0$ yields a partition
function that equals (\ref{CPdef}).

Because of this equivalence statistical physicists have shown increasing
interest in the chromatic polynomial, and in particular its zeros
in the complex plane. The original Lee-Yang theorem \cite{yanglee}
established the conditions on features of ferromagnetic
systems that ensure that all zeros of the partition function have
non-zero imaginary part, thereby bounding certain critical values
\cite{sokal01}. This kind of reasoning has been extended to antiferromagnetic
systems by what some authors refer to as {\em Lee-Yang theorems},
plural \cite{sokal00}. If in the $N\rightarrow\infty$ limit the
complex zeros of the partition function converge to pinch the real
axis this indicates the existence of singularities in the system at
these real values and so possible phase transitions or critical points.
For instance, the q-state Potts antiferromagnet may exhibit a critical
real $q_{c}>0$ above which the zero-temperature partition function
is analytic in the thermodynamic limit. Thus for $q<q_{c}$ the zero-temperature
phase is ordered, whereas the high-temperature phase ($T\rightarrow\infty$)
is generically disordered, indicating a phase transition in temperature
$T$ between ordered and disordered states. For $q>q_{c}$ the system
exhibits ground state disorder at $T\rightarrow0$, an exception to
the third law of thermodynamics \cite{huang}.

\begin{figure}[!t]
\begin{center}
\includegraphics[height=2in]{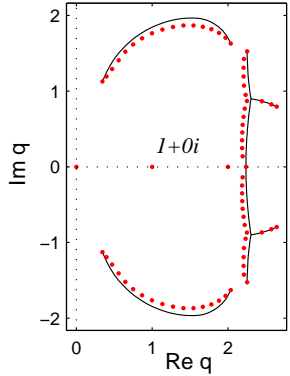}
\end{center}
\caption{Chromatic roots for strip lattices. Red points: roots for a $4 \times 18$
square lattice, free boundary conditions. Solid black line: limiting curve for a
$4 \times \infty$ lattice (again FBC).}
\label{rootschema}
\end{figure}

For computational reasons, the bulk of statistical physics research
on the chromatic polynomial has been conducted on strip lattices \cite{shrock01,shrock98,sokal01,shrock97}:
these are both sparse and have a repeating structure, and so are the
most computationally tractable graphs also for moderate sizes. This
structure lends itself to the use of specialized techniques, so lattices
are to date the largest graphs chromatic polynomial have been computed
for; in fact, for such strip graphs it is sometimes possible to calculate $N\rightarrow\infty$
limit sets of chromatic zeros analytically \cite{shrock01,sokal01}. Such lattice data represent
probably our most detailed knowledge of how the entire zero sets of
chromatic polynomials are laid out, and there are in fact certain
qualitative consistencies in these findings; figure \ref{rootschema}
gives an example of how the chromatic zero sets of strip lattices
are laid out. Most noticeable is that, for many lattice strips, zero sets invariably
trace out a somewhat elongated backwards `C' curling around
the point $1+0i$. Gaps and small horns are common features, showing
up in characteristic locations.

\section{Exploring the chromatic zeros of random graphs}
\begin{figure}[!t]
\begin{center}
\includegraphics[width=6.5in]{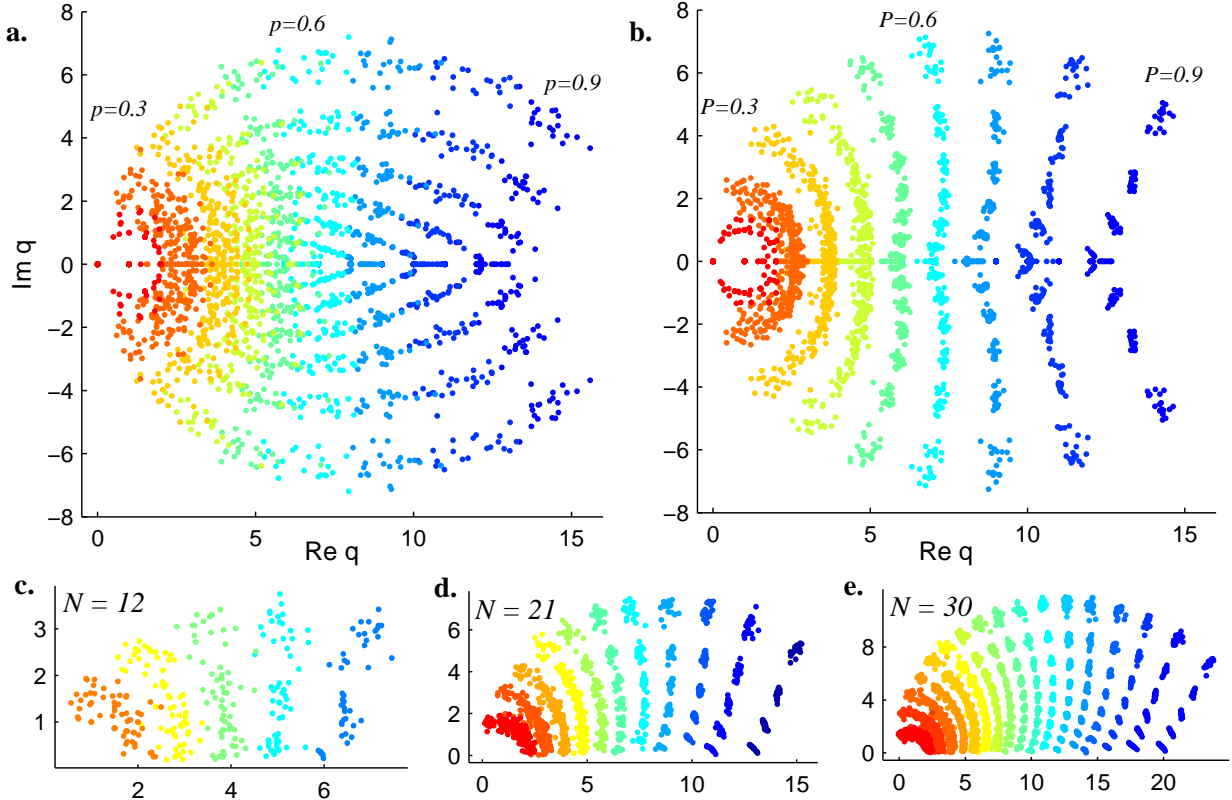}
\end{center}
\caption{Complex zeros of chromatic polynomials of random graphs. \textbf{(a)}
zeros for $G(20,p)$ with selection probability $p\in\{0.1,0.2,\dots,0.9\}$;
\textbf{(b)} zeros for $G(20,M)$ graphs with $M$ determined by fixed
edge density $P\in\{0.1,0.2,\dots,0.9\}$. \textbf{(c)}, \textbf{(d)},
and \textbf{(e)} zeros with positive real part for $G(N,M)$ graphs for
selected $N$, $M=\left[KN/2\right]$ determined by the average degree
$K\in\{3,4.5,6,\dots\}$. In all figures chromatic zeros for 20 sample graphs per
parameter set are plotted together. Density is indicated by colour,
running from red (low density) to blue (high density).}
\label{g20pall}
\end{figure}

To gain first insights toward chromatic zeros of random graphs, we
consider the two models of the ER random graph \cite{janson}: 
\begin{itemize}
\item [--] $G(N,p)$: the graph has $N$ vertices, between every pair
of vertices an edge exists with uniform probability $p$. The edge density $P$ of
any individual instance will then vary normally around $p$.
\item [--] $G(N,M)$: the graph is chosen uniformly from all graphs with
$N$ vertices and $M$ edges. The density $P$ then has the fixed value $M/\binom{N}{2}$.
\end{itemize}
In the $N\rightarrow\infty$ limit the two models are largely equivalent;
for finite $N$ they can give different results, but the main reasons
to use one or the other are practical. From the outset we had no clear
idea of what to expect, so we began with a preliminary investigation
of $G(N,p)$ graphs restricted to $N=20$. For each of the parameter
values $p\in\{0.1,0.2,\dots,0.9\}$ we generated 20 sample graphs;
we then calculated the chromatic polynomials with a \textsc{Form}
\cite{formp} implementation of the new algorithm \cite{timme} and
solved for the zeros of each polynomial numerically using \textsc{MPSolve}
\cite{mpsolv}. Complex zeros of all graphs plotted together are shown
in figure \ref{g20pall}(a). When first viewing these results we were
struck by the distinct banding in both the horizontal and vertical
directions, which implied that the complex zero sets could have a
very predictable layout depending on the basic graph parameters. After
confirming that similar patterns did not arise from zero sets of random
polynomials with similar coefficients we proceeded to the second phase
of our investigation, which was designed to measure the simultaneous
effect of both edge-density and the number of vertices on the zero
locations in a more systematic way.

\subsection{Parameters and constraints }
In these more extensive investigations we consider the $G(N,M)$ instead
of the $G(N,p)$ model. A cursory examination of
the $G(20,p)$ data showed that the positions of chromatic zeros with
non-zero imaginary part are very sensitive to the exact value of $P$,
For example, within each $p$ group we found the correlation between
the mean real value of these complex zeros and the exact $P$ value
(as measured from counting the edges in the generated graphs) was
between 0.982 and 0.994. Since we were interested in the effect
of density on zero location we then decided to directly control
for $P$. Figure \ref{g20pall}(b) shows zero sets for fixed-density
$G(N,M)$ graphs.

As well, we parameterized density by average degree $K = 2 M / N$
rather than directly by $P$, or by $M$. For a single value of $N$ it would
not matter which of these was used, since they can be simply converted one to another, but
over different $N$ the choice of metric affects the amount of data
generated and the ways one can aggregate it. A deciding factor was
that $K$ arises naturally in the study of physical systems
which feature a bounded number of interactions.

Finally, we added the constraint that graphs were to be 2-connected:
that is, every pair of vertices in the graph belong to at least one
cycle. ER random graphs experience a sharp connectivity transition
as density is increased \cite{janson}. Though the theoretical result
is in the large $N$ limit, the effect is manifest even for $N$ in
the range of the random graphs we used; none of our $G(20,0.1)$ graphs
were even 1-connected, while all of the $G(20,p)$ graphs for $p \geq 0.3$
were at least 2-connected. Since the chromatic polynomial of an unconnected
graph can be factorized into the chromatic polynomials of its connected
components, aggregating data from graphs without controlling connectivity
would have had the annoying effect of confounding the low-density
data for runs with different values of $N$, without otherwise telling
us anything we do not already know.

We used $N \in \{12,15,18,21,24,27,30\}$ and $K \in \{3,4.5,6,7.5,\dots,d\}$,
where $d$ is highest value divisible by 1.5 that is less than $N-2$.
For each of the 77 $(N,K)$ pairs 20 instances of $G(N,M)$ graphs
were generated using the integer $M$ that was closest to $KN/2$.
As above, chromatic polynomials were then calculated with the new
algorithm and zeros extracted using \textsc{MPSolve}. Figure \ref{g20pall}
(c), (d), and (e) show all chromatic zeros in the upper half complex
plane for some $N$. Note that the upper value and spacing of
the $N$ values chosen were primarily determined by computational
concerns; for low $N$ chromatic polynomials for all graphs could
be calculated in a matter of seconds, but in the high range a single
graph could require two or more days. In total 1540 random
graphs were generated, resulting in 35,700 individual data points
(chromatic zeros).

\subsection{Analysis of specific features of chromatic zeros}
From the combined data we extracted various graph invariants directly,
such as the chromatic number, number of real and integer zeros,
zero multiplicities, the maximum modulus, maximum real zero, and maximum
real and imaginary parts of zeros. Of particular interest to us were
the shape traced out by the complex chromatic zero set (i.e. zeros
with non-zero imaginary part) in the complex plane, and the point
at which this set approaches closest to the real axis.
The second of these we will denote as the {\em crossing point}
$X_{N,r}$ for the parameter $r$ (which can be average degree $K$
or density $P$ as required).

The crossing point is a finite analogue to the critical value $q_c$
discussed by previous authors \cite{shrock97,shrock98,shrock01},
which properly speaking can arise only in the $N\rightarrow\infty$
limit when critical phenomena arise, and the zero-set becomes continuous
such that its intersection with the real axis can be solved explicitly.
Since our graphs (and hence their chromatic zero sets) are finite
the crossing points $X_{N,r}$ were instead estimated in the following
way: 
\begin{itemize}
\item [--] For each parameter pair $(N,K)$ we collected the
chromatic zeros with positive imaginary value for all 20 graphs into
one set (zeros with negative imaginary value are their complex conjugates,
and can thus be neglected); 
\item [--] We fit an arc segment to each collected
zero set in the complex plane; moreover, we did a linear fit to determine
orientation (left or right leaning). All fits used least squares Euclidean
distance from each zero to the nearest point on the arc (or line)
as a metric. It should be noted here that centres and radii of arcs
were not constrained in any way (for example, to lie on the real axis). 
\item [--] The crossing point $X_{N,K}$ is then given by the positive
intersection of the arc and the real axis. As well, the curvatures
of the zero sets are obtained from the arc radii, and their orientations
from the slopes of the linear fits. 
\end{itemize}

\section{Scaling of chromatic zeros}
\begin{figure}[!b]
\begin{center}
\includegraphics[width=6in]{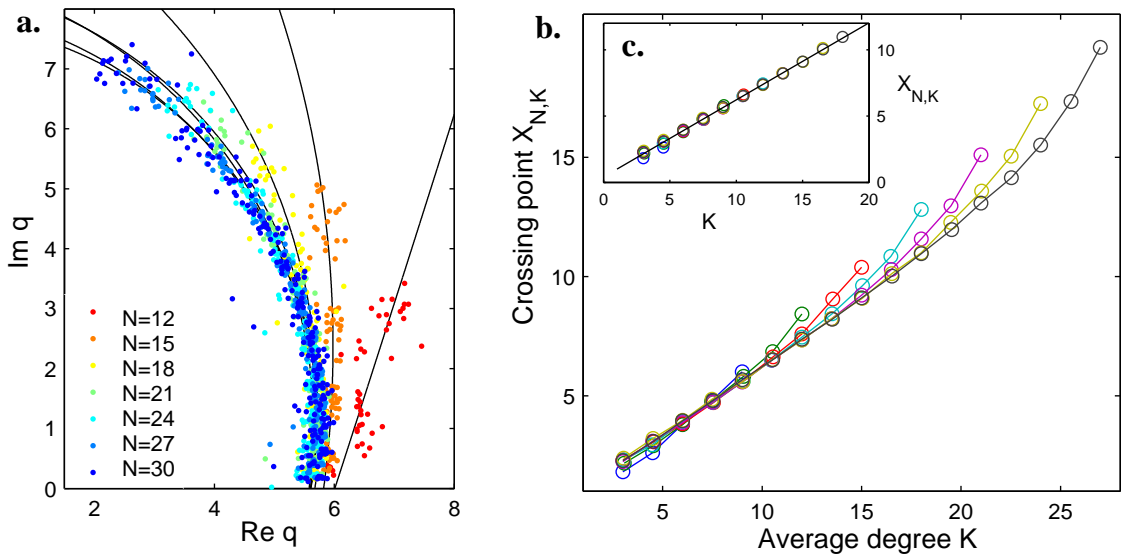}
\end{center}
\caption{Crossing points for chromatic zero sets. \textbf{(a)} Complex chromatic
zeros in upper half plane plus arc fits for average degree $K=9$,
various $N$. \textbf{(b)} Crossing point $X_{N,K}$ vs. $K$ for each
$N,K$. \textbf{(c)} (inset) Crossing points vs. average degree restricted
to graphs with density $P<.65$ with fit line $X_{N,K}= 0.58 K + 0.41$}
\label{fixedk}
\end{figure}
Preliminary visual inspection of the complex chromatic zeros as shown
in figure \ref{g20pall} revealed several consistent features. For
all $N$ tested, the zero sets at low densities are close to the origin
and laid out almost circularly around the point $1+0i$; as the density
is increased they gradually move away from the origin and uncurl until
they are almost vertical, after which the half-sets in the lower and
upper complex plane remained relatively straight, but become shorter
and angled, now away from the origin. 

Perhaps most surprising is the way that zero sets for graphs with
the same average degree $K$ almost always fell into roughly the same
location regardless of the $N$ value. Figure \ref{fixedk}(a) shows
complex zeros for $K = 9$. We note that the curvatures
of these zero sets do vary, and that those of the smallest (hence
densest) graphs pull to the right somewhat. Thus it would seem that
while location and therefore crossing point is largely determined
by average degree, features like the shape and height of the sets
will depend on $P$ and $N$ in a more complex way.

The zero locations (see figure \ref{fixedk}(b)) suggest that for
random graphs with edge-density below a certain cutoff $P_{c}$ the
crossing point $X_{N,K}$ scales linearly with $K$ but is otherwise
insensitive to changes of $N$. As we will show below, it also turns
out that $X_{N,K}$ has a predictable location for densities above $P_{c}$;
in this range, however, it is no longer a linear function of $K$
but a still relatively simple function of $N$ and $P$. We remark
that for low $K$ there is somewhat more variability in the data than
elsewhere.

\subsection{Determining cutoff density}
By most measures the cutoff density $P_{c}$ is somewhere between
0.6 and 0.7. Though by no means conclusive, the best empirical metrics
we have point to a value of $P_{c} \approx 0.65$. The most dispositive
evidence involves the shape and orientation of the zero
set. As we have noted, for a given $N$, as $K$ (or $P$) increases
the shape of the zero set consistently goes from an arc curving around
the origin to a pair of roughly straight lines leaning away from the
origin. Our working assumption is thus that the cutoff should occur
at the value of $K$ (or $P$) where the zero set is most vertical,
simultaneously consistent with either a very large arc curling left
or a straight line leaning imperceptibly right.
This can be made more precise by considering the slopes of
linear fits to the complex chromatic zero sets. For our data the slopes are negative for all
zero sets with $P \leq 0.648$, and positive for all zero sets with
$P \geq 0.652$ (no zero sets were associated with a density between
these values).

\subsection{Crossing point for $P<P_{c}$}

Crossing points for $P < P_{c}$ exhibit a close to linear scaling with
the average degree, cf. figure \ref{fixedk}(c). Fitting $X_{N,K}$ against $K$
below the cutoff gives the relation as
\begin{equation} \label{xnkltpc}
X_{N,K} = 0.582 K + 0.406.
\end{equation}
Goodness of fit tests
confirm that regardless of possible higher order terms the data points
are aligned very closely to the fit line. The reduced chi-squared
$\chi_{\textrm{r}}^{2}$ for the fit is a quite low 0.295 ($\chi_{\textrm{r}}^{2}$
is the standard $\chi^{2}$ normalized by the degrees of freedom;
a value of one or less indicates that the model accounts for the data
adequately or better, while values significantly greater than one
mean the fit is of poor quality).

We note that the variability in measured data is greater for low $K$
values than for high. This is in part due to the fact that we have
more low $K$ data points than high ones, and that variability of
zero location is definitely greater at low densities (noticeable from
simple visual inspection of zero plots). But there seems to be another
factor at work as well. For individual $N$ values the differences
between the calculated $X_{N,K}$ values and the fit are not scattered
about the line, but instead are larger (and consistently positive)
towards both ends of the applicable density range. As well, the crossing
point for $K=2$ (a cycle) should be at $Re(q)=2$, while the fit
line predicts a too-low
value of 1.57. This may imply that actual curves are either {\em
very} gently rounded, or that there is a separate low density regime
where a different slope applies. We do not currently have any good
criterion for distinguishing a second cutoff, however. Moreover, the
magnitude of these differences is small, and does not seem to increase
with larger $N$.

\subsection{Crossing point for $P>P_{c}$}
\begin{figure}[!b]
\begin{center}
\includegraphics[height=2in]{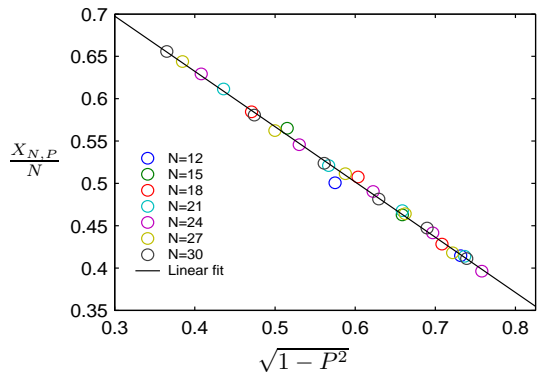}
\end{center}
\caption{Rescaled crossing points $X_{N,P}/N$ vs. $\sqrt{1-P^{2}}$, $P > P_{c}$}
\label{gtresc}
\end{figure}
The crossing points above the cutoff density do not scale directly
with $K$; in fact, it is not clear from visual inspection that they
conform to any simple function of $K$ or $P$. Hence it is something
of a surprise to find out that for $P > P_{c}$ the crossing points
are rather well described by
\begin{equation} \label{xnkgtpc}
X_{N,P} = N \left( a - b \sqrt{1 - P^{2}} \right),
\end{equation}
where $a \approx 0.9$ and $b \approx 0.65 \approx P_{c}$, as shown in
figure \ref{gtresc}. This relation was obtained by considering a
rescaling of the crossing point values to correct for the fact that the
distance from the cutoff crossing point $X_{N,P_{c}}$ to $N$ increases
as roughly $0.4N$ with $N$. It turned out that when rescaled this
way the $X_{N,P}$ values for different $N$ fell close to the same
arc segment when plotted against $P$; hence the $\sqrt{1-P^{2}}$
term in the relation above. The exact linear fit has a $\chi_{\textrm{r}}^{2}$
of 0.2868.

\subsection{Further quantities of interest}
\begin{figure}[!t]
\begin{center}
\includegraphics[height=2in]{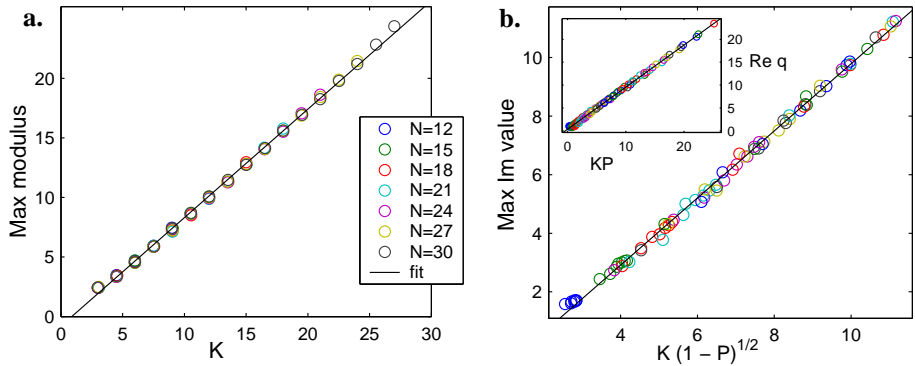}
\end{center}
\caption{\textbf{(a)} Maximum modulus (mean over each $N,K$ parameter set) vs. average degree $K$.
\textbf{(b)} Maximum imaginary value (mean over each $N,K$ parameter set) vs. $K \sqrt{1 - P}$. Inset
shows associated real values vs. $K P$.}
\label{maxmod}
\end{figure}

As mentioned above, we have considered a large number of other quantities
of interest. Each of these quantities requires a thorough scaling
analysis on its own, such that we here give only a brief summary of
the most important results and refer the reader to an extensive presentation
in a future publication. The maximum zero modulus $|z_{max}|$ and
the maximum imaginary value $\max(Im(z))$ both as well turn out
to have simpler scalings than were anticipated. The maximum modulus is well
approximated across all densities by the linear relation
\begin{equation} \label{maxmeq}
|z_{max}| = 0.91 K - 0.79
\end{equation}
though the exact relation again is probably non-linear at least in
the high and low density range; while all points are close to the
fit line the errors are not scattered normally but definitely pull
up at both ends (cf. figure \ref{maxmod}). The maximum imaginary value
does not grow linearly with $K$, but a linear relation can be obtained by rescaling: 
\begin{equation} \label{maximq}
\max(Im(z)) = 1.15 K \sqrt{1-P} - 1.67.
\end{equation}
A $\chi_{\textrm{r}}^{2}$ value of 0.296 again confirms a high quality fit.
Since $K = P (N-1)$, for fixed $P$ this relation does give a linear
scaling, this time in $N$. We note that $P \sqrt{1-P}$ achieves its
maximum at $P = \frac{2}{3}$; hence this result dovetails with our
finding of the density cutoff for the crossing point scaling at $P_{c} \approx 0.65$,
based on the most vertical layout of the chromatic zero set.

\section{Conclusions}
Applying the recently suggested vertex-based computational method
\cite{timme} yielded some corner-stone insights into the features
of chromatic polynomials of random graphs: the complex zeros of chromatic
polynomials of moderately sized random graphs systematically change
with the number of vertices $N$, the density $P$ and the average
degree $K$. For fixed $N$ the shapes of the complex zero sets evolve
as $P$ increases, from tracing an arc around the point $1+0i$ at
low $P$, to being almost vertically aligned at the cutoff density
$P_{c} \approx 0.65$, to sitting along pairs of straight lines leaning
away from the origin at higher $P$. The crossing point $X_{N,K}$
of the complex zeros scales linearly with the average degree $K$
for $P < P_{c}$; above this cutoff the crossing point appears to be
a simple function of $N$ and $\sqrt{1-P^{2}}$. If these results
continue to hold for larger $N$, the implications are straightforward.
Fixing the average degree $K$, we have $P \rightarrow 0$ as $N \rightarrow \infty$;
this suggests that for an infinite random graph with bounded degree
the chromatic zero set converges to an arc around $1+0i$ that crosses
the real axis at a finite position. On the other hand, if $P$ is
fixed or converges to a positive value then $K \rightarrow \infty$
and the crossing point and in fact all chromatic zeros with non-zero
imaginary value diverge to (complex) infinity. The former is most
relevant for statistical physics: if $G$ is an infinite disordered
system with bounded connectivity then in the zero-temperature limit
the zeros of antiferromagnetic Potts partition function meet the real
axis at a finite, predictable location that depends on the system's
coordination number.

Besides extending our studies to related polynomials such as Tutte
or reliability polynomials \cite{shrock03,tutte}, the
chromatic polynomial itself requires several future studies. In
particular, there is of course much that still needs to be done with
respect to Erd\"{o}s-R\'{e}nyi random graphs. Ideally we would like to obtain
results for larger $N$ for additional verification. However, while
the new method \cite{timme} does achieve a significant speedup, the
problem of computing the chromatic polynomial remains difficult and
so there is a limit to feasible $N,P$ values. 

Nevertheless, even for
accessible classes of graphs, several features of their chromatic
polynomials are not yet settled: While our results indicate that for
ER random graphs the location of chromatic zeros may be quite predictable,
there are many related graphs of interest with as yet unknown features.
Preliminary studies on two other classes of graphs, regular random
graphs and random bipartite graphs, suggest how far and in what direction
our current results might extend. As one might expect, with respect
to regular random graphs the main effect is to reduce the variation
in zero location across instances. The layout of the chromatic zero
sets for random bipartite graphs, on the other hand, is markedly different
than what we see for general ER random graphs, particularly at higher
densities. Since many lattices are as well bipartite this
bears closer study.

Two further `standard' classes of networks and graphs include: \\
 {\em ---Small-world graphs.} There is a possibility that our
scalings may depend on properties that will almost always hold for
small enough random graphs yet almost never hold for large enough
ones. One such property relates to the presence of triangles and other
small cycles. None of our graphs with average degree more than three
were triangle free, though for densities $P<1$ large enough random
graphs will usually be triangle free, and infinite random graphs are
locally indistinguishable from trees. Small-world graphs \cite{wattstrogatz},
which maintain their neighborhood properties
as they grow larger, are hence natural candidates for study. \\
 {\em ---Scale-free networks.} Some previous work on chromatic
polynomials has brought up the possibility that the presence of high
degree vertices may have a large influence on the location of some
of the graph's chromatic zeros \cite{sokal00}. We have not detected a pronounced
effect coming from the maximum degree itself but this might be due
to the strongly central, unimodal distribution of the degrees for
the random graphs we considered. Hence it is entirely possible that
the non-effect we noticed is due to our maximum degrees not being
large enough compared to the average degrees to exert an independent
influence. To test this supposition, a natural option would be to
compute chromatic zeros of graphs with scale-free degree distributions.
The vertex-based symbolic method we used \cite{timme} would need
to be adapted to compute reasonable size instances.

\section*{Acknowledgments}
We thank R. Ruppelt for a key hint. This work was supported by the
Max Planck Society via a grant to MT, by the German Ministry for Education
and Research (BMBF), grant no. 01GQ0430, and by the Max Planck Advisory
Committee for Electronic Data Processing (BAR).


\end{document}